\documentclass[useAMS,usenatbib]{mn2e}

\def\gsim{ \lower .75ex \hbox{$\sim$} \llap{\raise .27ex \hbox{$>$}} }
\def\lsim{ \lower .75ex \hbox{$\sim$} \llap{\raise .27ex \hbox{$<$}} }
\def\Mo{{\rm M_\odot}}

\title[Cusps in CDM halos]
{Cusps in CDM halos}

\author[J\"urg Diemand, Marcel Zemp, Ben Moore, Joachim Stadel,
$\&$ Marcella Carollo]
{J\"urg Diemand$^{1,3}$\thanks{diemand@physik.unizh.ch},
Marcel Zemp$^{1,2}$, Ben Moore$^{1}$, Joachim Stadel$^1$ $\&$ Marcella Carollo$^2$\\
$1$ Institute for Theoretical Physics, University of Z\"urich,
Winterthurerstrasse 190, CH-8057 Z\"urich, Switzerland\\
$2$ Institute of Astronomy, ETH Z\"urich,
ETH H\"onggerberg HPF D6, CH-8093 Z\"urich, Switzerland\\
$3$ Department of Astronomy and Astrophysics, University of California,
1156 High Street, Santa Cruz CA 95064, USA}
\begin{document}
%\date{Accepted 2005 ------. Received 2005 -----;
%in original form 2005 -----}

\pagerange{\pageref{firstpage}--\pageref{lastpage}}\pubyear{2005}
\maketitle
\label{firstpage}

\begin{abstract}
We resolve the inner region of a massive cluster forming in a
cosmological $\Lambda$CDM simulation with a mass resolution
of $2 \times 10^6 \Mo$ and before z=4.4 even $3 \times 10^5 \Mo$.
This is a billion times less than
the clusters final virial mass and a substantial increase
over current $\Lambda$CDM simulations.
We achieve this resolution using
a new multi-mass refinement procedure and
are now able to probe a dark matter halo density profile down 
to 0.1 percent of the virial radius. The inner
density profile of this cluster halo is well fitted by a power-law
$\rho \propto r^{-\gamma}$ down to the smallest resolved scale. 
An inner region with roughly constant logarithmic slope
is now resolved, which suggests that cuspy profiles
describe the inner profile better than recently
proposed profiles with a core. The cluster studied here
is one out of a sample of six high resolution
cluster simulations of \citet{Diemand2004pro} and
its inner slope of about $\gamma = 1.2$ lies
close to the sample average.
\end{abstract}


\section{Introduction}

Recently a great deal of effort has gone into high resolution simulations 
which have revealed density profiles of cold dark matter halos down
to scales well below one percent of the virial radius
(\citealt*{Fukushige2004}; \citealt*{Tasitsiomi2004}; 
\citealt*{Navarro2004}; \citealt*{Reed2004};
\citealt*{Diemand2004pro},''DMS04'' hereafter). But the form of
profile below $\sim 0.5$ percent of the virial radius 
remained unclear and there was no clear evidence for a cusp
in the center, i.e. no significant inner region with a 
constant logarithmic slope. Galaxy cluster halos would be
the ideal systems to resolve cusps numerically because 
of their low concentration. In a galaxy or dwarf halo the inner
power law is much harder to resolve because it lies at a smaller
radius relative to the size of the system.

The existence of a core or a cusp in the center of CDM halos
has important observational consequences and is 
the crucial point in many tests of the CDM theory.
Comparisons of dark matter simulations to rotation curves
of low surface brightness galaxies (LSB)
seem to favor constant density cores for most observed systems
(e.g. \citealt*{Moore1994}; \citealt*{Flores1994}; \citealt*{Salucci2000};
\citealt*{deBlock2001}; see, however \citealt*{vandenBosch2001};
\citealt*{Swaters2003}; \citealt*{Simon2004})x.
But these comparisons still depend to some extend on extrapolations
of the simulated profiles toward the center: \citet{Stoehr2004} 
extrapolate to a constant density core and claim
that the discrepancy to LSB galaxy rotation curves
is much smaller than previously believed.

The strength of the $\gamma$-ray signal from
dark matter annihilation depends on the square of the dark matter density
and the calculated flux values spread over
several orders of magnitude, depending on how one extrapolates the
density profiles from the known, resolved regions down into the centers
of the galactic halo and its subhalos (\citealt{Calcaneo2000}; 
\citealt{Stoehr2002}; \citealt{Bertone2005}; \citealt{Prada2004}).
Small, very abundant, Earth to Solar mass subhalos could be very luminous
in $\gamma$-rays if they are cuspy \citep{Diemand2005}.

The highest resolutions in cosmological simulations are reached with
the widely used refinement procedure 
(e.g. \citealt{Bertschinger2001}): 
First one runs a simulation
at uniform, low resolution and selects halos for re-simulation. Then one 
generates a new set of initial conditions using the same large scale fluctuations
and higher resolution and additional small scale fluctuations in the selected region.
With this technique \citet*{Navarro1996} were able 
to resolve many halos with a few ten thousand particles
and to infer their average density profile which 
asymptotes to an $\rho(r) \propto r^{-1}$ cusp. Other authors
used fitting functions with steeper (-1.5) cusps (\citealt*{Fukushige1997};
\citealt*{Moore1998}; \citealt*{Moore1999pro}; \citealt*{Ghigna2000}). 
Small mass CDM halos have higher
concentrations due to their earlier collapse \citep{Navarro1996}
but the slopes of the inner density profiles are independent
of halo mass (\citealt{Moore2001}; \citealt*{Colin2004}).
Open, ``standard'' and lambda CDM cosmologies, i.e.  
models with $(\Omega_M$,$\Omega_{\Lambda})=(0.3, 0.0)$,
$(1.0, 0.0)$ and $(0.3, 0.7)$ yield equal inner profiles 
(\citealt*{Fukushige2003}; \citealt*{Fukushige2004}).
There is some indication that models with less small scale power 
like WDM lead to shallower inner profiles 
(e.g. \citealt{Colin2000}; \citealt*{Reed2004}).
Different equation of states of the dark energy component 
lead to different collapse times and halo concentrations 
but it is not clear yet if it also affects slopes well inside of the
scale radius (\citealt{Maccio2004}; \citealt{Kuhlen2004}).
Most current simulations do not resolve a large enough
radial range to determine both the concentration and the inner
slope; at the current resolution these parameters show 
some degeneracy \citep{Klypin2001}.

Recently a large sample of $\Lambda$CDM
halos resolved with a million and more particles was simulated
(\citealt*{Springel2001cluster}; \citealt*{Tasitsiomi2004}; 
\citealt*{Navarro2004}; \citealt*{Reed2004};
\citealt*{Gao2005})
and the best resolved systems contain up to 25 million particles
(\citealt*{Fukushige2004}; DMS04). 
But even these very large, computationally expensive simulations resolved 
no inner region with a constant logarithmic
slope. (\citealt{Navarro2004};
\citealt{Stoehr2002}; \citealt{Stoehr2004}) 
introduced cored profiles which seem to 
fit the simulation data better than the cuspy profiles 
proposed earlier by \citet{Navarro1996} and 
\citet{Moore1999pro}. This better fit
was interpreted as indication against cuspy inner profiles. 
However these cored profiles have one
additional parameter and therefore it is not surprising
that they fit the data better.
DMS04 showed that an NFW-like profile with the inner
slope as additional free parameter fits the highest resolution profiles
just as well as cored profiles. Some theoretical arguments seem
to favor cusps (e.g. \citealt{Binney2004}; \citealt{Hansen2004}) 
but make only vague predictions about the inner slopes. 
A recent model combines simulation results and 
analytical arguments to predict an inner slope of -1.27 \citep{Ahn2005} 
At the moment higher resolution simulations
seem to be the only way to decide the core vs. cusp controversy. 

Here we present simulations of one of the galaxy clusters from
DMS04 with two orders of magnitude better mass resolution.
Our results give strong support to cuspy inner profiles. 
This increase in resolution was made possible 
with only a moderate increase in computational cost by 
using a new multi-mass refinement technique 
described in Section \ref{mmne}. In Section \ref{mmres} we present our results
and in Section \ref{mmconc} the conclusions.

\section{Numerical experiments}\label{mmne}

Table \ref{mmtab1} gives an overview of the simulations we present in this paper.
All runs discussed in this paper model the same 
$\Lambda$CDM cluster labeled ``D'' in DMS04.
With a mass resolution corresponding to $1.3\times 10^8$ and 
$1.04\times 10^9$ particles inside the virial radius of a cluster, DM25 and DM50 
are the highest resolution $\Lambda$CDM simulation performed so far.
Due to the large number of particles and the corresponding 
high force and time resolution these runs take a large amount of CPU time.
Fortunately the inner profiles of CDM clusters are already
in place around redshift one and evolve little between $z=4$ and $z=0$
(\citealt{Fukushige2004}; \citealt{Tasitsiomi2004}; \citealt{Reed2004}). 
Therefore one does not have to run the simulations to $z=0$ to gain insight
into the inner density profile. We stop DM50 at $z=4.4$, DM25 at $z=0.8$ 
and use the medium resolution runs D5 and D12 to quantify the low redshift
evolution of the density profile of the same cluster. 
Run DM25 was completed in about $2 \times 10^5$ CPU hours on the zBox 
supercomputer \footnote{http://www-theorie.physik.unizh.ch/$\sim$stadel/zBox/}.
The convergence radius of run DM50 is 1.7 kpc, estimated using 
the $r\propto N^{-1/3}$ scaling and the measured converged scales
from DMS04.

\begin{table*}
\centering
\begin{minipage}{140mm}
\caption{Parameters of the simulated cluster. At z=0 the viral mass is $3.1\times 10^{14} \Mo$ and
the virial radius is 1.75 Mpc. $N_{\rm HR}$ is the number of high resolution particles and
$m_{\rm HR}$ is the mass and $\epsilon_{\rm HR}$ the force softening length of these particles.
For the multi-mass runs we also give the masses ($m_{\rm LR}$) and softenings ($\epsilon_{\rm LR}$)
of the next heavier particle species. 
Softening lengths are given at z=0, ``[c]'' indicates that a constant softening
in comoving coordinates was used, ``[p]'' indicates that the softening was 
constant in physical units after z=9 and constant at ten times this value
in comoving units before z=9. The resolved scales are constant in physical units and
give the innermost radius we expect to resolve with the given mass resolution.
$N_{\rm vir,eff}$ is the actual number of particles within the virial radius at z=0 for
runs D6, D9 and D12. For the multi-mass runs it is the number needed to reach the same
resolution in the inner part by doing a conventional refinement of the entire system.
All runs are 300 Mpc cubes with periodic boundaries, well outside of the cluster forming 
region the resolution is decreased (as in DMS04).}
\label{mmtab1}
\begin{tabular}{l|c|c|c|c|c|c|c|c|c|c|c}
\hline
Run&$z_{start}$&$z_{end}$&$\epsilon_{\rm HR}$&$N_{\rm HR}$&$m_{\rm HR}$
&$\epsilon_{\rm LR}$&$m_{\rm LR}$&$r_{\rm resolved}$&$\eta$&time-&$N_{\rm vir,eff}$\\
 & & &[kpc]& &$[\Mo]$&[kpc]&$[\Mo]$&[kpc]& &step &\\
 \hline
 $D5$ & 52.4 &0&4.2[p]& 4'898'500&$3.0\;10^8$&-&-& 16.2&0.25&(\ref{ts})&$1.0\;10^6$\\

 $D6$ & 36.13 &0&3.6[p]& 31'922'181&$1.8\;10^8$&-&-& 13.5&0.2&(\ref{ts})&$1.8\;10^6$\\
 $DM6se$ & 36.13 &0&3.6[p]& 922'968&$1.8\; 10^8$&3.6[p]&$3.8\; 10^{10}$&13.5&0.2&(\ref{ts})&$1.8\;10^6$\\
 $DM6le$ & 36.13 &0&3.6[p]& 922'968&$1.8\; 10^8$&38.6[p]&$3.8\; 10^{10}$&13.5&0.2&(\ref{ts})&$1.8\;10^6$\\

 $D9$ & 40.27 &0&2.4[p]& 31'922'181&$5.2\;10^7$&-&-& 9.0&0.2&(\ref{ts})&$6.0\;10^6$\\
 $DM9$ & 40.27 &0&2.4[p]& 3'115'017&$5.2\; 10^7$&15[p]&$1.4\; 10^9$&9.0&0.2&(\ref{ts})&$6.0\;10^6$\\

 $D12$ & 43.31 &0&1.8[p]& 14'066'458 &$2.2\; 10^7$&-&-&6.8&0.2&(\ref{ts})&$1.4\;10^7$\\

 $DM25$ & 52.4 &0.8&0.84[c]& 65'984'375 &$2.4\; 10^6$&9[c]&$3.0\; 10^8$&3.3&0.25&(\ref{mmts})&$1.3\;10^8$\\
 $DM25lt$ & 52.4 &0.8&0.84[p]& 65'984'375 &$2.4\; 10^6$&9[p]&$3.0\; 10^8$&3.3&0.25&(\ref{ts})&$1.3\;10^8$\\

 $DM50$ & 59.3 &4.4 &0.36[c]& 16'125'000&$3.0\;10^5$&6[c]&$3.75\; 10^7$&1.7&0.25&(\ref{mmts})&$1.0\;10^9$\\

\end{tabular}
\end{minipage}
\end{table*} 

\subsection{Multi mass refinements}\label{mmref}

Often in cosmological N-body simulations one uses high resolution particles only where 
one halo forms and heavier particles in the surroundings to account
for the external tidal forces. One usually tries to defines a large enough
high resolution region to minimize or avoid mixing
of different mass particles within the region of interest.
One exception is \citet{Binney2002} who used particles of two
different masses everywhere to estimate the amount of 
two body relaxation in cosmological simulations.
In plasma simulations on the other hand multi mass
simulations have been successfully used since the 1970s
(e.g. \citealt{Dawson1984} and refs. therein). Here we apply this
idea to increase the resolution in the core of one cluster halo in a 
cosmological N-body simulation.

The refinement procedure is usually applied to entire virialised systems, i.e. one
marks all particles inside the virial radius of the selected halo and traces
them back to the initial conditions.
Then one refines the region that encloses the positions of the marked particles.
Usually the region is further increased to prevent any mixing of low resolution
particles into the virial radius of the final system. In DMS04
all particles within 4 comoving Mpc in the initial conditions were added to the
high resolution region. This assures that only light particles end up
within the virial radius of the final cluster and it also has the advantage
that halos in the outskirts of the cluster (out to 2 or 3 virial radii) are still
well resolved \citep{Moore2004}. But with this procedure only between
one fourth to one third of all the high resolution particles end up
in the cluster.

If one is only interested in the inner regions of a halo it is possible to use
a new, more efficient way of refinement: Instead of refining
the whole virialised system we only refine the region where the inner particles
come from. This allows to reduce the size of the high resolution region 
considerable, because most of particles that end up near the center of the system 
start in a very small region, compared to the region which one finds by tracing back all the
particles inside the virial radius. Using this technique it is possible to reduce the
computational cost of a CDM cluster simulation by at least one order of magnitude at equal force and mass
resolution in the inner region. Of course now one has different mass particles inside
the final virialised structure therefore we must verify that significant 
equipartition and relaxation (\citealt{Binney2002};\citealt{Diemand2004rel})   
is not occurring and affecting the final results.
In section \ref{proof} we show that the density profiles
of such multi-mass clusters (runs DM6le and DM9) are the same as 
the ones of fully refined clusters at equal 
peak resolution (runs D6 and D9).

In this paper we apply the multi mass refinement to the cluster 'D' from DMS04.
This cluster is well relaxed and isolated at 
z=0 and has an average density profile and inner slope close to the mean value.
First we mark all particles within one percent of the virial radius in the final halo
and trace them back to the initial conditions. Then we add all particles within one
comoving Mpc of a marked particle to the set of marked particles, 
and finally we add all particles which
lie on intersections of any two already marked particles
on the unperturbed initial grid positions.
After these two steps there is region with a fairly regular triaxial boundary which
contains only marked particles. The number of marked particles grows by almost
a factor of 8 during these additions, but it is still more than a factor of two
smaller than the number of particles in the final cluster and a factor
of ten smaller than the original high resolution 
volume used in DMS04. The computational cost
with our code and parameters is roughly proportional to the number of 
high resolution particles, therefore we gain about a factor of
ten with this reduction of the high resolution region.
Probably one can reduce the high resolution volume further and
focus even more of the computational effort into the innermost region, 
we plan to explore this possibility with future simulations.

\subsection{Code and parameters}

The simulations have been performed using 
PKDGRAV, written by Joachim Stadel and Thomas Quinn \citep{Stadel2001}
using the same cosmological 
and numerical parameters as in DMS04
with a few changes given below and in Table \ref{mmtab1}.
The cosmological parameters are $(\Omega_m,\Omega_{\Lambda},\sigma_8,h)
=(0.268, 0.732, 0.7, 0.71)$. The value of $\sigma_8=0.9$
given in DMS04 is not correct:
During the completion of this paper we found that
due to a mistake in the normalization
our initial conditions have less power than intended.
This lowers the typical formation redshifts and halo concentrations slightly
but does not affect the slopes of the inner density profiles.

We use the GRAFICS2 package \citep{Bertschinger2001} to
generate the initial conditions.
The particle time-step criterion $\Delta t_i < \eta\sqrt{\epsilon/a_i}$
, where $a_i$ is the acceleration of particle ``i'',
gives almost constant time-steps in the inner regions of a halo
(see Figure 2 in DMS04), but the
dynamical times decrease all the way down to the center.  
Therefore the time-step criterion was slightly modified, to make sure
enough time-steps are taken also near the halo centers:
Instead of 
\begin{equation} \label{ts}
\Delta t_i < \eta\sqrt{\epsilon/a_i}
\end{equation}
we now use
\begin{equation} \label{mmts}
\Delta t <\min(\eta\sqrt{\epsilon/a_i}, \eta / 4\sqrt{G\rho_i} )  \;\; ,
\end{equation}
where $\rho_i$ is the density at the position of particle ``i'', 
obtained by smoothing over 64 nearest neighbors.
We used $\eta = 0.25$ for runs DM25 and DM50. 
Note that in the
inner region of a CDM halo $\rho(r) \simeq 0.6 \rho(< r)$,
i.e.  0.8 $\sqrt{G\rho(r_i)} \simeq \sqrt{G\rho(< r_i)}$ 
therefore the condition (\ref{mmts}) with $\eta = 0.25$ assures that
at least 12 time-steps per local dynamical time $1/ \sqrt{G\rho(< r_i)}$ 
are taken. 

The time-steps are obtained by dividing the main time-step $(t_0/200)$
by a factor of two until condition (\ref{mmts}) is fulfilled.
In runs DM25 and DM50 the smallest particle time-steps are $t_0/51200$.
According to Figure 2 in DMS04 this time-step is sufficient to 
resolve smaller scales than 0.1 percent of the virial radius, 
i.e. less than the limit set by the mass resolution, even in run DM50.

\begin{figure}
\vskip 3 truein
\includegraphics{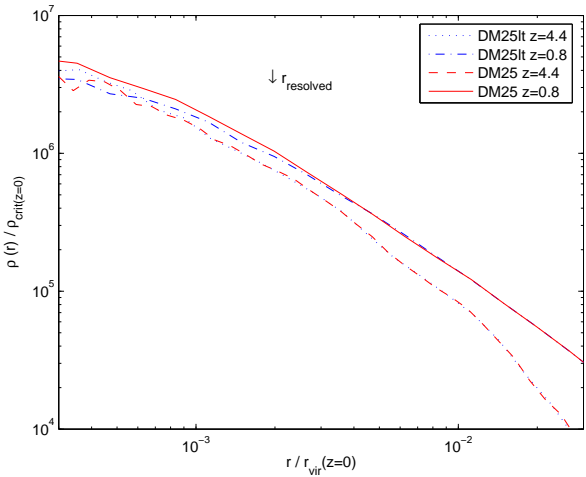}
\caption{\label{plotMultiTimestep.eps} 
Density profiles in physical (not comoving) coordinates
at redshifts 4.4 and 0.8. The two runs have equal mass resolution
but different time-steps and softening. The arrow indicates the
resolution limit set by the particle mass. The run with the larger
time-steps and softening underestimates the dark matter density
outside of the resolution scale.}
\end{figure}

The smaller time-steps in the inner regions of the cluster are
crucial: In Figure \ref{plotMultiTimestep.eps} we compare two runs 
which only differ in the time-step criterion. DM25lt was run with the
standard criterion (\ref{ts}) and $\eta=0.2$, for run DM25
we used the more stringent, computationally more expensive 
criterion (\ref{mmts}) and $\eta=0.25$. The
difference in CPU time is about a factor of two. At z=0.8
the densities in run DM25lt are clearly lower 
out to 0.003 virial radii which also affects part of the region
we aim to resolve with this run ($r_{\rm resolved}= 0.0019 r_{\rm vir}$).
Due to the high computational cost of these runs we cannot perform 
a complete series of convergence test at this high resolution
but due to the monotonic convergence behavior of PKDGRAV 
for shorter time-steps \citep{Power2003} we are confident
that DM25 is a better approximation to the true CDM density profile
of this cluster.

Our time-stepping test confirms that the time resolution in DMS04 was sufficient
to resolve the minimum scale of 0.3\% virial radii set by their mass resolution.
For the purpose of this work, i.e. to resolve a region even closer to the center
smaller time-steps are necessary.
These two runs illustrate nicely how a numerical parameter or criterion
that passes convergence tests performed at low or medium resolution can 
introduce substantial errors if employed in high resolution runs.

\subsection{Testing the multi mass technique}\label{proof}

Reducing the high resolution region in the way described in Section \ref{mmref}
produces multi mass virialised systems, i.e. halos where particles of different mass
are mixed up with each other. The inner regions are dominated by light particles
and the region near the virial radius by heavier particles. But one will find
particles of both species everywhere in the final halo and one has to worry
if this mixing introduces numerical effects, like energy transfer from the
outer part to the inner part (from the heavy to the light particles)
due to two body interactions. This could lead to numerical flattening of
the density profile and make heavy particles sink to the center 
(\citealt{Binney2002}; \citealt{Diemand2004rel}). 

\begin{figure}
\vskip 3 truein
\includegraphics{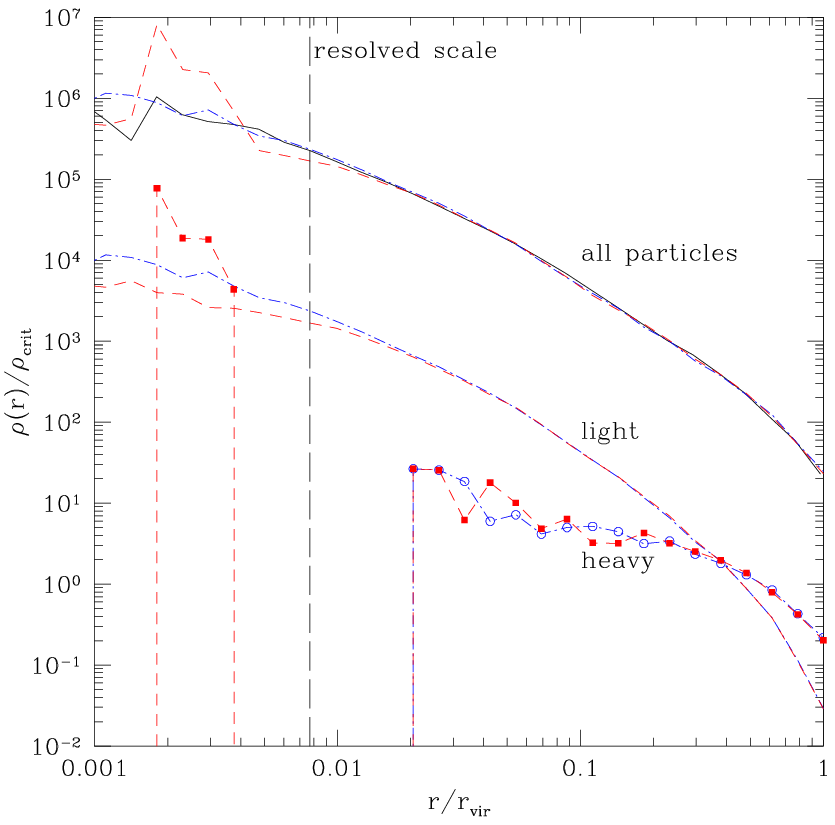}
\caption{\label{D6pro0.eps} 
Tests of the multi-mass refinement technique. The upper three lines shows
the total density profile at z=0 from the fully refined run
D6 (solid lines) and the multi-mass runs DM6se (dashed) 
and DM6le (dashed dotted). The lower lines 
(same line styles, offset by two magnitudes for clarity)
show the density profiles of the two particle species, 
i.e. of the light ones (lines without symbols) 
and of the heavier ones 
(lines with symbols: filled squares for D6se, open circles for D6le).
The vertical dashed line indicates the innermost resolved scale.
In the multi-mass run with more softened heavier particles (D6lh) the
inner profile is dominated by light particles and identical to the
fully refined run of the same cluster (D6). When the heavier particles
have short softenings some of them spiral into the center 
due to dynamical friction and transfer
heat to the light particles. This affects the total density profile,
i.e. it is lower near the resolved scale and has a bump due to
a condensation of cold, massive particles very close to the center.
}
\end{figure}

\begin{figure*}
\vskip 5.5 truein
\includegraphics{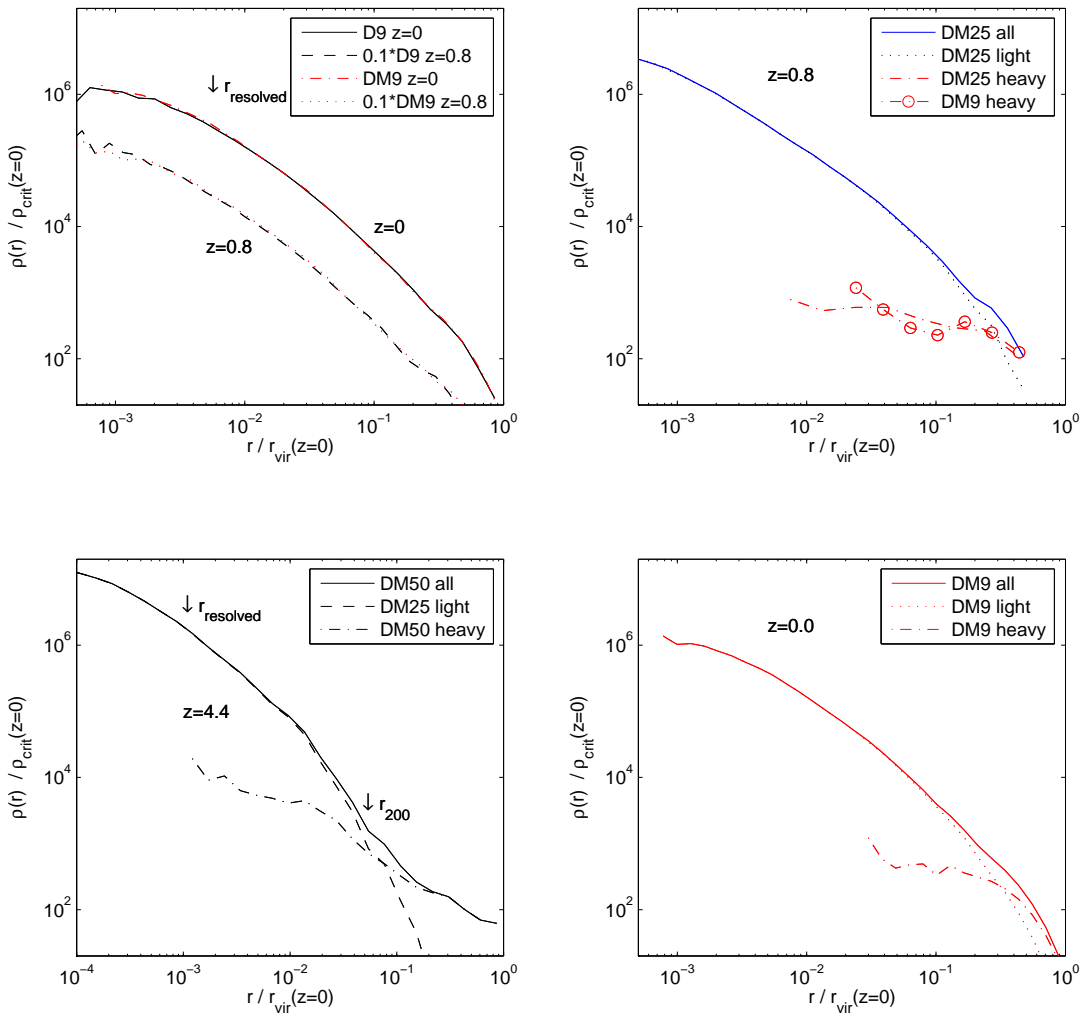}
\caption{\label{plotMultiCONVpro.eps} 
Tests of multi-mass refinement and convergence. The upper left panel shows
that run D9 which contains only high resolution particles within
the virial radius has the same density profile as the
multi-mass run DM9. The z=0.8 profiles are shifted downward
by a factor of ten for clarity. The arrows indicate the convergence radius
of run D9 estimated in DMS04.
The lower left panel shows the high and low resolution particles
in run DM50 at z=4.4.
The panels on the right illustrate the mixing of light 
and heavy particles in runs DM9 and DM25 
which have the same refinement regions.}
\end{figure*}

To check if the multi mass technique works for
cosmological simulations we re-ran the simulations
D6 and D9 from DMS04 using a reduced high resolution region.
We call these multi-mass runs ``DM6se'', ``DM6le'' and ``DM9'' (see Table \ref{mmtab1}).
The next heavier particles in the surrounding region
are 216 times more massive in DM6se and DM6le 
and 27 times more massive in DM9. The heavier particles 
in DM6le and DM9 have
larger softening to suppress discreteness effects
while DM6se uses the same small softening for both species. 
Figure \ref{plotMultiCONVpro.eps}
shows that the density profiles of the fully refined run D9 and
the partially refined run DM9 are identical over the entire
resolved range. Figure \ref{D6pro0.eps} shows that the same is true for
run DM6le, the larger mass ratio of 216 does not introduce any
deviation form the density profile of the fully refined run.

A small softening in the heavier species (run DM6sl) does introduce
errors in the final density profile (Figure \ref{D6pro0.eps}).
The total mass profile
is shallower near the resolved radius and has a high density
bump below the resolved scale. The light particles are more extended
and the bump is caused by a cold, dense condensation of six heavy
particles within 0.004 $r_{\rm vir}$. These six heavy particles
have a 3D velocity dispersion of only 273 km/s, while the light
particles in the same region are much hotter, $\sigma_{3D}=926$ km/s.
They are hotter than the particles in the same region in run
D6 and DM6le (both have only light particles in this inner part),
the dispersion are 722 km/s for D6 and 708 km/s for DM6le.

These tests indicate that the reduced refinement regions work well
in runs D9M and DM6le and therefore we used the same 
refinement regions to set up the higher resolution run
DM25. In this run the heavier particles are 125 times more massive than the
high resolution particles and they have a softening of 9 kpc. 
For run DM50 we refined only the inner part of the most massive cluster
progenitor at z=4.4 in the same way as the final cluster in runs DM6le, DM6se, DM9 and DM25.
In run DM50 the heavier particles are also 125 times more massive than the
high resolution particles. 

Figure \ref{plotMultiCONVpro.eps} shows 
how the initially separated
species of light and heavy particles mix up during the 
the runs DM9, DM25 and DM50.
The density profiles profiles of DM6le and
DM9 do not suffer from numerical effects due to the multi-mass 
setup. This indicates that the same is true for run DM25 which has
the same refinement regions. In run DM50 the 
amount and location of mixing at z=4.4 relative to $r_{200}$ is very similar to
the situation if DM9 at z=0.0, therefore we expect DM50 to 
have the same density profile as a fully refined cluster, i.e. as a
cluster resolved with a billion particles. 

\section{The inner density profiles} \label{mmres}

Here we try to answer the question if the inner density profiles
of dark matter halos have a constant density or 
a cusp $\rho(r) \propto r^{-\gamma}$. 
At resolutions of up to 25 million particles within 
the virial radius there is no evident convergence 
toward any constant inner slope 
(\citealt{Fukushige2004}; DMS04).

\subsection{Results of run DM25}

\begin{figure}
\vskip 3 truein
\includegraphics{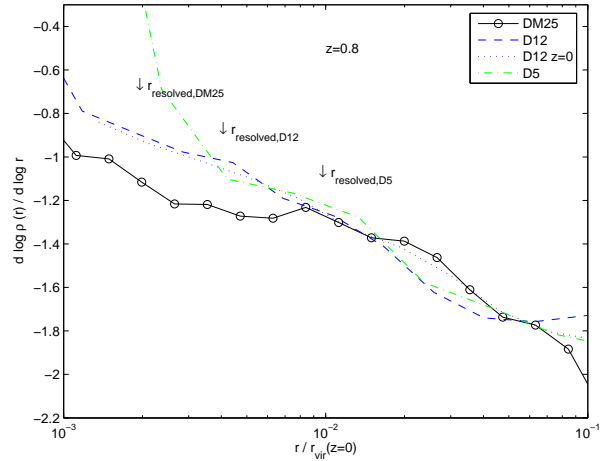}
\caption{\label{plotMultiSlopeStep100.eps} 
Logarithmic slope of the density profile of run DM25 at z=0.8. The slopes
of runs D5 and D12 at z=0.8 and z=0 are also shown for comparison. 
The arrows indicates the estimated convergence radii. Note that although
the densities at the converged scales are within 10 percent the
density gradients can already be substantially smaller.}
\end{figure}

Run DM25 has an effective resolution corresponding to 127 million particles within 
the virial radius and a force resolution of $0.48 \times 10^{-3} r_{\rm vir}$.
At this up to now unmatched resolution the inner slope is
roughly constant from the resolved radius 
(see Figure \ref{plotMultiSlopeStep100.eps}) out to about one percent
of the virial radius of the final cluster.

Run D12 resolves the same cluster with 14 million particles and
shows no convergence to a constant inner slope. Note that the ``D'' 
cluster is one of six clusters analyzed in DMS04 and its inner
profile is not special and rather close to the sample average. 

Figure \ref{plotMultiSlopeStep100.eps} indicates that there is a cusp in the 
centers of cold dark matter clusters and it becomes apparent only
at this very high numerical resolution. The non-constant slopes just
near the convergence scale are probably due to the first signs of
numerical flattening that set in at this scale. At higher densities below
the resolved scales one cannot make any robust predictions yet, but if one has
to extrapolate into this region Figure \ref{plotMultiSlopeStep100.eps} motivates
the choice of a cusp $\rho(r) \propto r^{-\gamma}$ with $\gamma\simeq 1.2$.

\subsection{Resolving the very inner density profile at z=4.4 (run DM50)}

Mass accretion histories show that the inner part of CDM halos
is assembled in an early phase of fast accretion 
(\citealt{vandenBosch2002}; \citealt{Wechsler2002}; \citealt{Zhao2003})
and recent high resolution simulations revealed that the inner density profile does
not evolve at low redshift
(\citealt*{Fukushige2004}; \citealt*{Tasitsiomi2004}; \citealt*{Reed2004}).
Figure \ref{plotMultiSlopeStep100.eps} confirms that the inner density profile
of runs D12 and D5 do not change from z=0.8 to z=0.

Therefore in run DM50 we focus our computational effort even more on the early evolution of 
the inner profile. We refine the inner region of the most massive progenitor identified 
in run DM25 at z=4.4. Since the refinement region needed is much smaller than the one
of DM9 or DM25 and we only run the simulation to z=4.4 it is feasible to go to a much
better mass and force resolution. The high resolution particles in run DM50 are a billion
times lighter than the final cluster.

\begin{figure}
\vskip 3 truein
\includegraphics{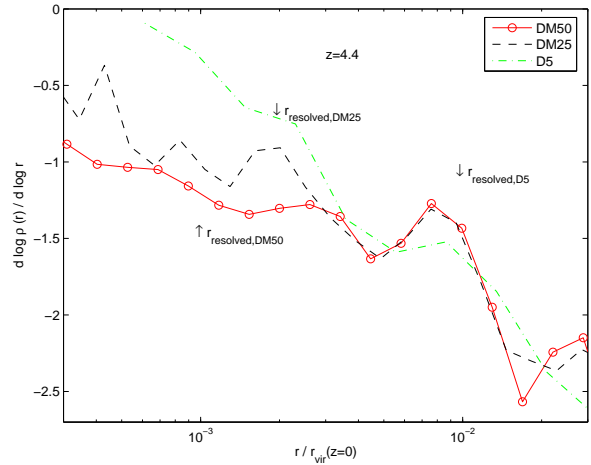}
\caption{\label{plotMultiSlopeStep20.eps} 
Logarithmic slope of the density profile of run D5, DM25 and DM50 at z=4.4.
The arrows indicates the estimated convergence radii. A constant inner
slope of about -1.2 is evident in the highest resolution run DM50. The
increase of the slopes around the resolved radii is due to the onset of numerical
flattening.}
\end{figure}

Figure \ref{plotMultiSlopeStep20.eps} shows that the density profile of run DM50
at z=4.4 is cuspy down to the resolved radius (0.1 \% of the final virial radius). As in
run DM25 the slopes begin to shallow just at the converged scale due to 
numerical flattening.
The profile of DM50 at z=4.4 supports the finding from run DM25 that the inner profile
follows a steep power law $\rho \propto r^{-1.2}$. At the higher resolution of run DM50
we find substantially higher physical densities in the cluster center at z=4.4 compared
to lower resolution runs like DM25.
This suggests that a run like DM50 evolved to low redshift
would also yield substantially higher central densities
as currently resolved in the centers of runs like D12 and DM25.

\subsubsection{Estimating the z=0 profile of a billion particle halo}

\begin{figure}
\vskip 3 truein
\includegraphics{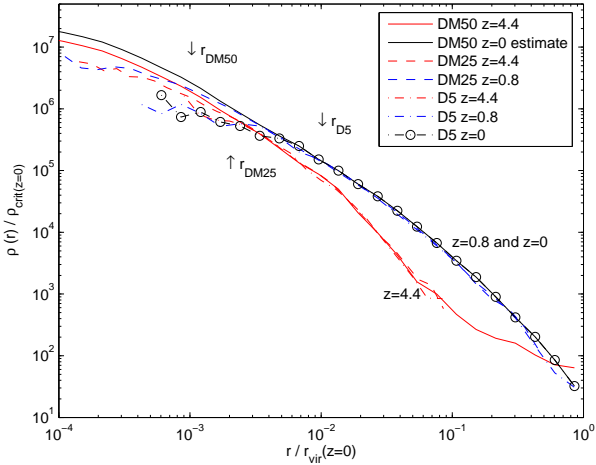}
\caption{\label{plotMultiProEv.eps} 
Density profiles in physical (not comoving) coordinates
at redshifts 4.4, 0.8 and 0. Arrows mark the resolved scales
of each run. The densities in the inner part do not evolve between
z=0.8 and z=0 and the inner slopes remain constant even from 
z=4.4 to z=0. Using these observations we are able to estimate the
final profile of a billion particle halo (upper solid line).}
\end{figure}

Now we go one step further and use the information from all the ``D''-series runs to 
try to estimate
the density profile one would obtain if one simulates this cluster with a billion particle
all the way to present time, a run which would be possible but
extremely expensive with today's computational resources. From Figure \ref{plotMultiProEv.eps}
one finds that the density profile of run DM25 near its resolution scale shifts upward by a
constant factor of 1.4 from z=4.4 to z=0.8. The density around 0.01 $r_{\rm vir,z=0}$ is
constant form z=0.8 to z=0, see run D5 in Figure \ref{plotMultiProEv.eps}. 
The inner density profile {\it slopes} are constant even longer,
i.e. from z=4.4 to z=0, see Figures \ref{plotMultiSlopeStep100.eps} and \ref{plotMultiSlopeStep20.eps}.
Therefore we estimate the z=0 profile of run DM50 by rescaling the z=4.4
profile of DM50 by a factor 1.4 and using
the z=0 profile of run D12 outside of 0.005 $r_{\rm vir,z=0}$
(see Figure \ref{plotMultiProEv.eps}).
The extrapolated z=0 profile of run DM50 should be regarded as a best guess for 
the density profile of an average CDM cluster resolved with a billion particles.
A (multi-mass) simulation with this (effective) resolution evolved to
redshift zero would be needed to
check the accuracy of the estimate performed here. Note that our conclusions are
based on the $z=0.8$ results from run DM25 and not on the somewhat
uncertain $z=0$ extrapolation proposed
in this section (but they are consistent with it). 

\subsection{Inner slope estimates based on the enclosed mass}

\begin{figure}
\vskip 3 truein
\includegraphics{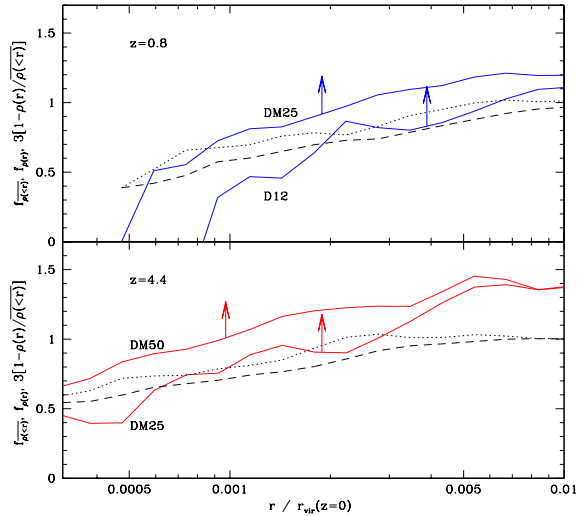}
\caption{\label{maxslope.eps} 
$\gamma^*(r)$ for the
two highest resolution runs at z=4.4 and z=0.8 (solid lines)
and fractions of the densities of these two runs 
(dotted lines for $\rho(r)$ and dashed lines for $\bar{\rho}(<r)$).
Due to different convergence rates in local and cumulative quantities
$\gamma^*(r)$ values from the lower resolution runs lie below
the higher resolution results in the inner part of the halo.
The arrows at $r_{\rm resolved}$ correct for this effect based on the following observations:
The ratio $\rho(r_{\rm resolved})/\bar{\rho}(<r_{\rm resolved})$ is typically underestimated
(0.87 of the high resolution value) due to a small deficit in local density (0.95 of the true value)
and a larger one in the enclosed density
(0.83 of the true value) due to missing mass in the innermost regions.
Underestimating $\rho(r_{\rm resolved})/\bar{\rho}(<r_{\rm resolved})$
by 0.87 leads to $\gamma^*$ values which are to small by about 0.3.
}
\end{figure}

For a mass distribution which follows $\rho(r) \propto r^{-\gamma}$ all the way 
in to $r=0$ the slope $\gamma$ can be calculated at any radius using the local density and
the mean enclosed density \citep{Navarro2004}: 
$\gamma^*(r) = 3 ( 1 - \rho(r)/\bar{\rho}(<r) )$. For simulated CDM density
profiles where $\gamma$ becomes
smaller towards the center $\gamma^*(r)$ is an upper limit for the asymptotic inner
slope as long as {\it both} $\rho(r)$ and $\bar{\rho}(<r)$ have fully converged
at radius $r$. Convergence tests show that
the enclosed density $\bar{\rho}(<r)$ converges slower than the local density
and $\bar{\rho}(<r)$ is generally underestimated near $r_{\rm resolved}$ due 
to missing mass within $r_{\rm resolved}$ (\citealt{Power2003}; DMS04).
Figure \ref{maxslope.eps} shows $\gamma^*(r)$ for the
two highest resolution runs available at z=4.4 and z=0.8. We also plot the fractions 
of the local densities of the two runs and the fractions of enclosed densities 
to illustrate the different convergence scales of local and cumulative quantities.
Figure \ref{maxslope.eps} confirms that at the estimated resolved scales for 
D12 and DM25 the local densities are within 10\% of the higher resolution runs
\footnote{We determine $r_{\rm resolved}$ by demanding that the local density
has to be within ten percent of the value from a much higher resolution run
and in cases where no such run is available the measured convergence radii
form lower resolution runs are rescaled using the mean inter-particle separation
$r_{\rm resolved} \propto N_{\rm vir}^{-1/3}$ (see DMS04).}.
The typical differences are even smaller (about 5\%). The enclosed density
$\bar{\rho}(<r)$ however converges slower: At $r_{\rm resolved}$ we find
that the values are only about 0.83 of those measured
in the higher resolution runs. This causes the ratio
$\rho(r_{\rm resolved})/\bar{\rho}(<r_{\rm resolved})$ to be underestimated
(about 0.87 of the true value). This propagates into a larger relative error in 
$\gamma^*(r_{\rm resolved})$ which turns out to be too low by about 0.3 for the
profiles studied here (given the arrows at $r_{\rm resolved}$ in Figure \ref{maxslope.eps}).
The different convergence rates of local and cumulative quantities tend to
produce artificially low $\gamma^*(r)$ values and this effect becomes especially 
large near $r_{\rm resolved}$. The significance of $\gamma^*(r)$ appears to be difficult
to interpret, but the convergence tests presented here and in DMS04 
suggest that $\gamma^*(r_{\rm resolved})$
is not a robust upper limit for the asymptotic inner slope.

\subsection{Cored and cuspy fitting functions}

In this section we fit one cuspy and two recently proposed cored functions
to the density profiles of DM25 at z=0.8 and to the tentative z=0 extraploation
from run DM50. From the last section we expect
the cuspy function to work better in the inner part but we try to fit
also the cored profiles for comparison.

We use a general $\alpha\beta\gamma$-profile
that asymptotes to a central cusp $\rho(r) \propto r^{-\gamma}$:  
\begin{equation}\label{mmGpro}
\rho_{\rm G}(r) = \frac{\rho_s}{(r/r_s)^{\gamma}(1 + (r/r_s)^{\alpha})^{(\beta-
\gamma)/\alpha} } \,.
\end{equation}
If one takes $\alpha$, $\beta$ and $\gamma$ as free
parameter one encounters strong degeneracies, 
i.e. very different combinations of parameter
values can fit a typical density profile equally well \citep{Klypin2001}.
Therefore we fix the outer slope $\beta = 3$ 
and the turnover parameter $\alpha = 1$. 
For comparison the NFW profile has $(\alpha, \beta, \gamma) = (1,3,1)$, the M99 
profile has $(\alpha, \beta, \gamma) = (1.5,3,1.5)$. We fit the three parameters
$\gamma$, $r_s$ and $\rho_s$ to the data.

\citet{Navarro2004} proposed a different fitting function which curves smoothly over 
to a constant density at small radii:
\begin{equation}\label{mmNpro}
\ln(\rho_{\rm N}(r)/\rho_s) = 
(-2/\alpha_{\rm N}) \left[ (r/r_s)^{\alpha_{\rm N}}  - 1 \right] \,
\end{equation}
$\alpha_{\rm N}$ determines how fast this profile 
turns away from a power law in the inner part. \cite{Navarro2004} found that
$\alpha_{\rm N}$ is independent of halo mass and
$\alpha_{\rm N} = 0.172 \pm 0.032$ for all their simulations, including 
galaxy and dwarf halos.

Another profile that also curves away from power law behavior 
in the inner part was
proposed by \citet{Stoehr2002}:
\begin{eqnarray}
\rho_{\rm SWTS}(r) = \frac{V_{max}^2}{4 \pi G} \ 
10^{-2 a_{\rm SWTS} \left[\log\left(\frac{r}{r_{max}}\right)\right]^2} 
\frac{1}{r^2} \times \nonumber \\
\times \ \left[1-4 \ a \ \log\left(\frac{r}{r_{max}}\right) \right]
\end{eqnarray}
where $V_{max}$ is the peak value of the circular velocity,
 $r_{max}$ is the radius of the peak and $a_{\rm SWTS}$ 
determines how fast the profile turns away from an power 
law near the center.
\cite{Stoehr2004} found that cluster profiles are well
fitted with this formula using  $a_{\rm SWTS}$ values between 0.093
and 0.15. 

\begin{figure}
\vskip 3 truein
\includegraphics{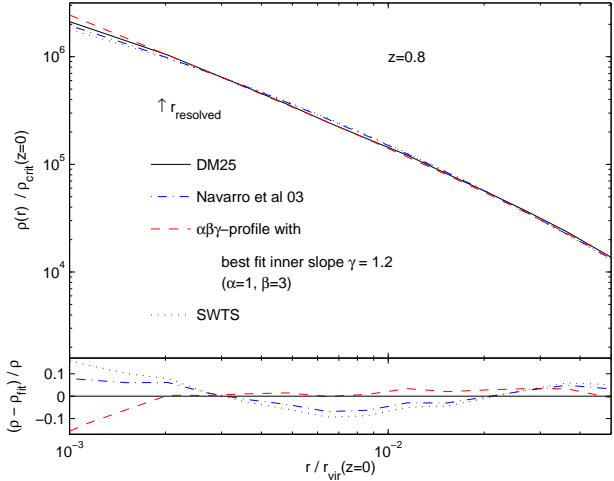}
\caption{\label{plotMultiPro2ParamFitsStep100.eps} 
Density profile of run DM25 at z=0.8 and fits with three different functions.}
\end{figure}

\begin{figure}
\vskip 3 truein
\includegraphics{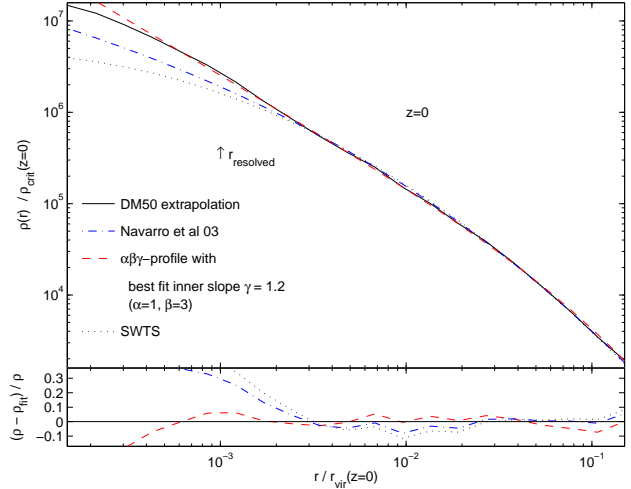}
\caption{\label{plotMultiPro2ParamFits.eps} 
Density profile of run DM50 extrapolated to z=0.0
and fits with three different functions.}
\end{figure}

These three functions were fitted to the data from z=0.8 
by minimizing the relative
density differences in each of about 20 logarithmically spaced bins in
the range resolved by DM25 (i.e. form $0.0019 r_{\rm vir,z=0}$ = 3.3 kpc to 
 $r_{\rm vir,z=0}$ = 1750 kpc). At z=0 we use the resolved range of D12
for the fits (i.e. form $0.0039 r_{\rm vir,z=0}$ = 6.8 kpc to 
 $r_{\rm vir,z=0}$). The resulting best fit values and
the root mean squares of the relative density differences are given in Table \ref{mmfp}.

\begin{table*}
\centering
\begin{minipage}{140mm}
\caption{Density profile parameters of run DM25 at z=0.8 and of DM50 extrapolated to z=0. 
$\Delta$ is the root mean square of $(\rho-\rho_{\rm fit})/\rho$
for the three fitting functions used.}
\label{mmfp}
\begin{tabular}{l | c | c | c || c | c | c || c | c | c }
  \hline
redshift& $\gamma_{\rm G}$&$r_{\rm sG}$[kpc]&$\Delta_{\rm G}$&
 $\alpha_{\rm N}$&$r_{\rm s\;N}$ [kpc]&$\Delta_{\rm N}$&
 $a_{\rm SWTS}$&$r_{\rm max\;SWTS}$[kpc]&$\Delta_{\rm SWTS}$  \\
  \hline
  0.8 & 1.20 & 260 & 0.075 & 0.157 & 233 & 0.076 & 0.130 & 565 & 0.087\\
  0.0 & 1.20 & 283 & 0.059 & 0.162 & 236 & 0.133 & 0.140 & 518 & 0.179\\
  \hline   
\end{tabular}
\end{minipage}
\end{table*}

At z=0.8 the average residuals of the three fits are similar, but they are dominated
by the contribution from the outer parts of the cluster (see Figure 6 in DMS04).
Figures \ref{plotMultiPro2ParamFitsStep100.eps} 
and \ref{plotMultiPro2ParamFits.eps}
show that in the inner part the cuspy
profile describes the data better.
Both cored profiles {\it underestimate} the measured density at the 
resolution limit both at z=0.8 and in the
estimated z=0 profile. These profiles lie {\it below} the 
measured density profiles even {\it inside} of $r_{\rm resolved}$ where one has to expect 
that the next generation of simulations will be able to resolve 
even higher densities.

\begin{figure}
\vskip 3 truein
\includegraphics{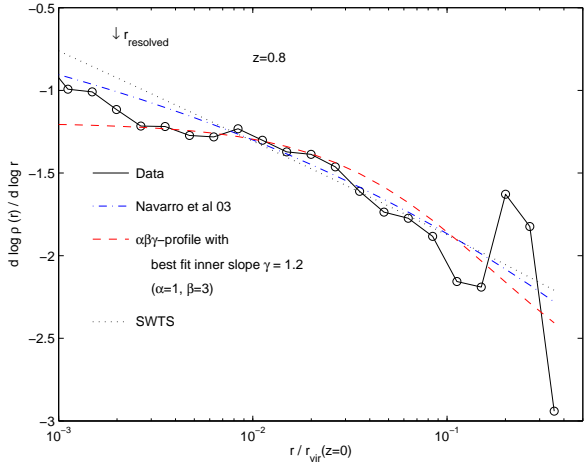}
\caption{\label{plotMultiSlope2ParamFitsStep100.eps} 
Logarithmic slopes of the measured and fitted 
density profiles from Figure \ref{plotMultiPro2ParamFitsStep100.eps}.}
\end{figure}

\begin{figure}
\vskip 3 truein
\includegraphics{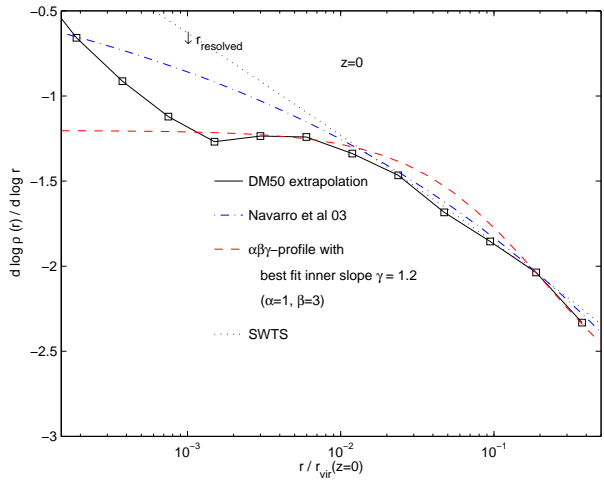}
\caption{\label{plotMultiSlope2ParamFits.eps} 
Logarithmic slope of the extrapolated z=0 DM50 density profile and of the fitted 
density profiles from Figure \ref{plotMultiPro2ParamFits.eps}.}
\end{figure}

Figures \ref{plotMultiSlope2ParamFitsStep100.eps} 
and \ref{plotMultiSlope2ParamFits.eps} show the slopes of
the simulated profile in comparison with the slopes of the best fits.
Again it is evident that in the inner part the cuspy profile
describes the real density run better.

\section{Conclusions} \label{mmconc}

The main conclusions of this work are the following:

\begin{itemize}

\item It is possible to use different mass particles
to resolve one halo in cosmological CDM simulations
without affecting the resulting density profiles.

\item This ``multi-mass'' technique allows a reduction 
of the necessary number of particles and the computational
cost by at least one order of magnitude without loss of
resolution in the central region of the halo.

\item We confirm that the inner profile of a typical CDM cluster
does not evolve since about redshift one.

\item The logarithmic slope of the dark matter density
profile converges to a roughly constant value in the inner part
of cluster halos. This probably holds also for smaller systems 
(like galaxy and dwarf halos) but there it is  even more
difficult to numerically resolve the cusps.

\item At resolutions around 10 million particles per halo
the inner slope appears to approach zero continuously but
this impression is caused by numerical flattening
of the profiles due to insufficient mass resolution.

\item The cluster studied here has a central cusp
$\rho \propto r^{-\gamma}$ with a slope of about 
 $\gamma = 1.2$. From earlier studies (DMS04) 
we expect this inner profile to be close
to the average and the scatter is about $0.15$.

\item Profiles with a core (\citealt{Stoehr2002};\citealt{Navarro2004}) 
{\it underestimate} the measured dark matter
density at (and even {\it inside} of) the current resolution limit.

\end{itemize} 

\section*{Acknowledgments}

We like to thank Piero Madau and Mike Kuhlen
for useful discussions
and the referee for valuable suggestions.
All computations were performed on the zBox
supercomputer at the University of Zurich.
J. D. is supported by the Swiss National Science Foundation.

\bsp
\label{lastpage}
\end{document}